\documentclass[aps,preprintnumbers,floatfix,article,amsmath,amssymb,floatfix,10pt,prd,superscriptaddress,nofootinbib,scrartcl,twocolumn]{revtex4-2}
\usepackage{bm}
\usepackage{amsfonts}
\usepackage{latexsym}
\usepackage[latin1]{inputenc}
\usepackage{graphicx}
\usepackage{amsmath}
\usepackage{palatino}
\usepackage{mathpazo}
\usepackage[british]{babel}
\usepackage{hhline}
\usepackage{multirow}
\usepackage{textcomp}
\usepackage{float}
\usepackage{booktabs}
\usepackage{dcolumn}
\usepackage{hhline}
\usepackage{multirow}
\usepackage{ragged2e}
\usepackage{hyperref}
\hypersetup{colorlinks,citecolor=blue}
\hypersetup{colorlinks=true,linkcolor=red,filecolor=magenta,    urlcolor=cyan}
\usepackage{amsmath}
\usepackage{xcolor}
\usepackage{orcidlink}
\usepackage{epsfig}
\usepackage{caption}
\usepackage{subcaption}
\usepackage{commath}
\def\jnl@style{\it}
\def\aaref@jnl#1{{\jnl@style#1}}

\def\aaref@jnl#1{{\jnl@style#1}}

\def\aj{\aaref@jnl{AJ}}                   
\def\apj{\aaref@jnl{ApJ}}                 
\def\apjl{\aaref@jnl{ApJ}}                
\def\apjs{\aaref@jnl{ApJS}}               
\def\apss{\aaref@jnl{Ap\&SS}}             
\def\aap{\aaref@jnl{A\&A}}                
\def\aapr{\aaref@jnl{A\&A~Rev.}}          
\def\aaps{\aaref@jnl{A\&AS}}              
\def\mnras{\aaref@jnl{Mon.~Not.~Roy.~Astron.~Soc.}}             
\def\prd{\aaref@jnl{Phys.~Rev.~D}}        
\def\prc{\aaref@jnl{Phys.~Rev.~C}}  
\def\prl{\aaref@jnl{Phys.~Rev.~Lett.}}    
\def\qjras{\aaref@jnl{QJRAS}}             
\def\skytel{\aaref@jnl{S\&T}}             
\def\ssr{\aaref@jnl{Space~Sci.~Rev.}}     
\def\zap{\aaref@jnl{ZAp}}                 
\def\nat{\aaref@jnl{Nature}}              
\def\aplett{\aaref@jnl{Astrophys.~Lett.}} 
\def\apspr{\aaref@jnl{Astrophys.~Space~Phys.~Res.}} 
\def\physrep{\aaref@jnl{Phys.~Rep.}}      
\def\physscr{\aaref@jnl{Phys.~Scr}}       
\def\commat{\aaref@jnl{Comm.~Math.~Phys.}}              
\def\science{\aaref@jnl{Science}}               
\def\cqg{\aaref@jnl{Classical Quant.~Grav.}}            
\def\jpcs{\aaref@jnl{JPCS}}                                     
\def\ijmpd{\aaref@jnl{Int.~J.~Mod.~Phys.~D}}                    
\def\grg{\aaref@jnl{Gen.~Relat.~Gravit.}}               
\def\rpp{\aaref@jnl{Rep.~Prog.~Phys.}}          
\def\npa{\aaref@jnl{Nucl.~Phys.~A}}        
\def\lrr{\aaref@jnl{Living Rev.~Rel.}}                   
\def\jcap{\aaref@jnl{J.~Cosmology Astropart.~Phys.}}    
\def\rmp{\aaref@jnl{Rev.~Mod.~Phys.}}   
\def\epjc{\aaref@jnl{Eur.~Phys.~J.~C}} 
\def\plb{\aaref@jnl{~Phy.~Lett.~B}} 
\def\mpla{\aaref@jnl{Mod.~Phy.~Lett.~A}} 
\def\arxiv{\aaref@jnl{arxiv.org}}

\allowdisplaybreaks[1]

\begin{document}
\color{black}       
\title{\bf Exploring cosmological evolution and constraints in $f(T)$ teleparallel gravity}

\author{M. Koussour\orcidlink{0000-0002-4188-0572}}
\email[Email: ]{pr.mouhssine@gmail.com}
\affiliation{Department of Physics, University of Hassan II Casablanca, Morocco.} 

\author{A. Altaibayeva}
\email[Email: ]{aziza.ltaibayeva@gmail.com \textcolor{black}{(Corresponding author)}}
\affiliation{Department of General and Theoretical Physics, L.N. Gumilyov Eurasian National University, Astana 010008, Kazakhstan.}

\author{S. Bekov}
\email[Email: ]{ss.bekov@gmail.com}
\affiliation{Department of General and Theoretical Physics, L.N. Gumilyov Eurasian National University, Astana 010008, Kazakhstan.}
\affiliation{Kozybayev University, Petropavlovsk, 150000, Kazakhstan.}

\author{F. Holmurodov}
\email[Email: ]{hfarhod2024@gmail.com}
\affiliation{Institute of Fundamental and Applied Research, National Research University TIIAME, Kori Niyoziy 39, Tashkent 100000, Uzbekistan.}
\affiliation{Faculty of Mathematics, Namangan State University, Boburshoh str. 161, Namangan 160107, Uzbekistan.}

\author{S. Muminov\orcidlink{0000-0003-2471-4836}}
\email[Email: ]{sokhibjan.muminov@gmail.com}
\affiliation{Mamun University, Bolkhovuz Street 2, Khiva 220900, Uzbekistan.}

\author{J. Rayimbaev\orcidlink{0000-0001-9293-1838}}
\email[Email: ]{javlon@astrin.uz}
\affiliation{Institute of Fundamental and Applied Research, National Research University TIIAME, Kori Niyoziy 39, Tashkent 100000, Uzbekistan.}
\affiliation{University of Tashkent for Applied Sciences, Str. Gavhar 1, Tashkent 100149, Uzbekistan.}
\affiliation{Urgench State University, Kh. Alimjan Str. 14, Urgench 221100, Uzbekistan}
\affiliation{Shahrisabz State Pedagogical Institute, Shahrisabz Str. 10, Shahrisabz 181301, Uzbekistan.}


\begin{abstract}

This study explores the extension of teleparallel gravity within the framework of general relativity, introducing an algebraic function $f(T)$ dependent on the torsion scalar $T$. Motivated by the teleparallel formulation, we investigate cosmological implications, employing the simplest parametrization of the dark energy equation of state. Our chosen $f(T)$ function, $f(T)=\alpha(-T)^n$, undergoes stringent constraints using recent observational data ($H(z)$, SNeIa, BAO, and CMB). The model aligns well with cosmic dynamics, exhibiting quintessence behavior. The evolution of the deceleration parameter, the behavior of dark energy components, and the $Om(z)$ diagnostic further reveal intriguing cosmological phenomena, emphasizing the model's compatibility with quintessence scenarios.\\

\end{abstract}

\maketitle


\section{Introduction}

This verification of late-time acceleration, primarily driven by observations of Type Ia Supernovae (SNeIa) \cite{Riess/1998,Perl/1999}, Baryon Acoustic Oscillations (BAOs) \cite{Eisenstein/2005, Percival/2007}, Cosmic Microwave Background (CMB) \cite{Komatsu/2011}, and Hubble parameter measurements $H(z)$ \cite{Farooq/2017}, has sparked extensive research efforts aimed at unraveling its underlying mechanisms. To elucidate late-time acceleration, various models have been proposed, with dark energy (DE) emerging as a successful candidate connected to the cosmological constant. While the conventional DE models have provided valuable insights, exploring alternative avenues involves modifying the geometry beyond general relativity. Some notable approaches include $f(R)$ gravity (where $R$ represents the curvature scalar) \cite{Staro/2007,Capo/2008,Chiba/2007}, coupling matter and curvature in $f(R,\mathcal{T})$ gravity (where $\mathcal{T}$ represents the trace of the energy-momentum tensor) \cite{Harko/2011,Moraes/2017}, $f(Q)$ gravity (where $Q$ represents the non-metricity scalar) \cite{Jim/2018}, and $f(R,G)$ gravity (where $G$ represents the Gauss-Bonnet scalar) \cite{Laurentis/2015,Gomez/2012}. These theories aim to provide alternative explanations for the DE phenomenon by incorporating modifications in the curvature.

To explain the behavior of the universe at cosmic scales, researchers have explored more general geometries beyond the Riemannian geometry, which is applicable at the solar system level. Teleparallel Gravity (TG), a form of modified gravity that employs a torsion scalar instead of a curvature scalar, has gained increasing attention. In TG, the metric of spacetime is replaced by a set of tetrad vectors, serving as the physical variables that describe gravitational properties. This approach uses the Weitzen$\ddot{o}$ck connection, offering a distinct mathematical framework. The extension of the action of modified gravity based on torsion leads to a unique class of modified gravity known as the Teleparallel Equivalent of General Relativity (TEGR) or $f(T)$ gravity. Numerous studies in $f(T)$ gravity have explored cosmological solutions \cite{Paliathanasis/2016}, thermodynamics \cite{Salako/2013}, late-time acceleration \cite{Myrzakulov/2011, Bamba/2011}, cosmological perturbations \cite{Chen/2011}, large-scale structure \cite{Li/2011}, cosmography \cite{Capozziello/2011}, energy conditions \cite{Liu/2012}, matter bounce cosmology \cite{Cai/2011}, wormholes \cite{Jamil/2013}, anisotropic universe \cite{Rodrigues/2016,Koussour/2022}, and observational constraints \cite{Nunes/2016}. Recently, Zhadyranova et al. \cite{Zhadyranova} explored late-time cosmic acceleration through a detailed investigation of a linear $f(T)$ cosmological model, using observational data. In addition, a comprehensive analysis of $f(T)$ gravity can be found in \cite{Cai/2016}.

Furthermore, exploring models beyond the cosmological constant is crucial for a comprehensive understanding of the expansion of the universe. One effective approach involves parameterizing the Equation of State (EoS) of DE, which describes the relationship between pressure and energy density. In the standard cosmological model, the EoS parameter is assumed to be constant with a value of $\omega_{\Lambda}=-1$ for the cosmological constant. However, allowing the EoS to vary over cosmic time provides deeper insights into the underlying physics of DE. The CPL parametrization \cite{w-CPL1,w-CPL2} is a valuable two-parameter model that captures deviations from a constant EoS value in DE studies. Beyond CPL, more sophisticated parametrizations are available, including the JBP parametrization \cite{w-JBP}, logarithmic parametrization \cite{w-LOG}, and BA parametrization \cite{w-BA}. These parameterizations provide flexibility in exploring DE scenarios beyond the cosmological constant. Each parametrization comes with unique features, capable of capturing different aspects of DE behavior. Moreover, they offer a versatile framework for testing diverse DE models and comparing their predictions with observational data. This parametric approach enhances our ability to discern subtle variations in the behavior of DE and contributes to a more nuanced understanding of the dynamic universe.

The primary objective of this study is to delve into the extended $f(T)$ modified gravity, commencing from the TEGR as opposed to GR. In this investigation, we adopt the simplest parametrization for the EoS parameter and derive exact solutions for the modified Friedmann equations within the framework of the FLRW spacetime. Specifically, we employ the power-law model of $f(T)$ gravity, expressed as $f(T)=\alpha(-T)^n$, where $\alpha$ and $n$ are model parameters. Subsequently, we rigorously constrain these parameters utilizing a comprehensive set of observational datasets encompassing Hubble parameter measurements $H(z)$, SNeIa, and BAO. Our particular focus is on scrutinizing the late-time accelerating behavior of the $f(T)$ gravity model through the lens of various cosmological parameters.

The structure of this work unfolds as follows: We commence in Sec. \ref{sec2} by providing an overview of TG and subsequently introduce the framework of $f(T)$ gravity. Sec. \ref{sec3} delves into the cosmological aspects of the model, presenting solutions to the field equations. The treatment of observational data and the methodological approach for constraining the relevant parameters are detailed in Sec. \ref{sec4}. A thorough examination of the late-time accelerated phase is undertaken in section \ref{sec5}, focusing on the dynamics behavior of cosmological parameters. In Sec. \ref{sec6}, we employ the $Om(z)$ diagnostic test to distinguish our cosmological model from other DE models. The work culminates in Sec. \ref{sec7} with a comprehensive conclusion.

\section{Field equations in $f(T)$ theory} \label{sec2}

The $f(T)$ theory of gravity, which is modified based on the torsion scalar, features a geometric action determined by an algebraic function associated with the torsion. Analogously to TG, this theory employs orthonormal tetrad components defined within the tangent space at every point of the manifold to articulate the geometric elements. Generally, the line element can be expressed as such:
\begin{equation}
    ds^2=g_{\mu \nu} dx^{\mu} dx^{\nu}=\eta_{ij} \theta^{i} \theta^{j}.
\end{equation}

Here, we establish the following components:
\begin{equation}
    dx^{\mu}=e_{i}^{\,\, \mu}\theta^{i}, \quad \theta^{i}=e^{i}_{\,\, \mu}dx^{\mu},
\end{equation}
where $\eta_{ij} = \text{diag}(1, -1, -1, -1)$ represents the metric associated with Minkowskian spacetime, and $\{e^{i}_{\,\,\mu}\}$ denotes the components of the tetrad satisfying the conditions:
\begin{equation}
e_{i}\,\,^{\mu} e^{i}\,\,_{\nu} = \delta^\mu_\nu, \quad e_{\mu}\,\,^{i} e^{\mu}\,\,_{j} = \delta^i_j. \label{eq:identity}
\end{equation}

The connection utilized in this theory follows $Weitzenb\ddot{o}ck$ prescription \cite{Aldrovandi/2013},
\begin{equation}
\Gamma^{\alpha}_{\mu \nu}= e_{i}^{\,\, \alpha} \partial_{\mu} e^{i}_{\,\,\nu}=- e^{i}_{\,\,\mu} \partial_{\nu} e_{i}^{\,\, \alpha}\label{eq:connection}
\end{equation}

With this connection, the torsion tensor's components are expressed as,
\begin{equation}
T^{\alpha}_{\,\,\mu \nu}= \Gamma^{\alpha}_{\,\,\nu \mu}- \Gamma^{\,\,\alpha}_{\mu \nu}= e_{i}^{\,\,\alpha}\left(\partial_{\mu} e^{i}_{\,\,\nu}-\partial_{\nu} e^{i}_{\,\,\mu} \right).
\label{eq:torsion}
\end{equation}

This tensor contributes to the definition of the contorsion tensor:
\begin{equation}
K^{\mu \nu}_{\,\, \alpha} = -\frac{1}{2}\left(T^{\mu \nu}_{\,\, \alpha} - T^{\nu \mu}_{\alpha} - T_{\alpha}^{\,\, \mu \nu}   \right),
\label{eq:contorsion}
\end{equation}

These objects, torsion, and contorsion, combine to form the tensor $S_{\alpha}^{\,\, \mu \nu}$, 
\begin{equation}
    S_{\alpha}^{\,\, \mu \nu} = \frac{1}{2}\left(K^{\mu \nu}_{\,\, \alpha} + \delta^{\mu}_{\alpha} T^{\lambda \mu}_{\,\, \lambda}- \delta^{\nu}_{\alpha} T^{\lambda \mu}_{\,\, \lambda}   \right).
    \label{eq:tensorS}
\end{equation}

The torsion scalar, denoted as $T$, is defined using the tensor $S_{\alpha}^{\,\, \mu \nu}$ and the torsion tensor \cite{Maluf/2013,Cai/2016},
\begin{equation}
T = S_{\alpha}^{\,\, \mu \nu} T^{\alpha}_{\,\,\mu \nu}= \frac{1}{2} T^{\alpha \mu\nu} T_{\alpha \mu\nu} + \frac{1}{2} T^{\alpha \mu\nu} T_{ \nu\mu \alpha} - T_{\alpha \mu}^{\,\, \,\, \alpha} T^{\nu \mu}_{\,\,\,\, \nu} .\label{eq:torsion_scalar}
\end{equation}

The action for the modified $f(T)$ theory with matter is then expressed as:
\begin{equation}
\label{7}
    S =\frac{1}{2 \kappa^2} \int d^4x\, e[T+f(T)]+\int d^4x\, e\, \mathcal{L}_m,
\end{equation}
where $e$ represents the determinant of the tetrad, given by $e=det(e^{i}_{\,\,\mu})=\sqrt{-g}$, where $g$ is the determinant of the space-time metric. In addition, $f(T)$ denotes an algebraic function dependent on the torsion scalar $T$ and $\mathcal{L}_{m}$ is the matter Lagrangian. 

Taking the variation of the action (\ref{7}) with respect to the tetrads leads to the field equations,
\begin{align} \label{1d}
& S_{\mu}^{\,\, \nu \rho} \partial_{\rho}Tf_{TT} + \left[ e^{-1}e_{\mu}^{i}\partial_{\rho}\left(ee_{i}^{\,\, \mu}S_{\alpha}^{\,\, \nu \lambda}\right) + T_{\,\, \lambda \mu}^{\alpha}S_{\alpha}^{\,\, \nu \lambda}\right] f_{T} \nonumber \\
& + \frac{1}{4} \delta_{\mu}^{\nu}f = \frac{\kappa^{2}}{2}\mathcal{T}_{\mu}^{\nu},
\end{align}
where $f_T={\partial f}/{\partial T} $, $ f_{T T}={\partial^2 f}/{\partial T^2}$, and $\mathcal{T}_{\mu}^{\nu}$ is the energy-momentum tensor defined as,
\begin{equation}
\mathcal{T}_{\mu}^{\nu}=(\rho+p)u_\mu u^\nu - p \delta_{\mu}^{\nu},
\end{equation}
where $\rho$ and $p$ denote the energy density and pressure, respectively, of the ordinary matter comprising the universe. The four-velocity, denoted as $u^\mu$, satisfies the condition $u^{\mu} u_{\nu}=1$. 

In this work, we consider the flat FLRW metric, which, as is customary, facilitates the application of the aforementioned theory within a cosmological framework, leading to the derivation of modified Friedmann equations. The flat FLRW metric is expressed as
\begin{equation}
\label{9}
 ds^{2}=dt^{2}-a^{2}(t) \delta_{ij}dx^{i} dx^{j}, 
\end{equation}
where $a(t)$ represents the scalar factor of the universe. Consequently, evaluating the torsion scalar yields:
\begin{equation}
T=-6 H^2.
\end{equation}

By analyzing Eqs. (\ref{1d}) to (\ref{9}), we derive the modified Friedmann equations as follows:
\begin{align}
& 6H^2 + 12H^2 f_T + f = 2\kappa^2 \rho, \label{F1} & \\
& 2\left(2\dot{H} + 3H^2\right) + f + 4\left(\dot{H} + 3H^2 \right) f_T \nonumber \\
& - 48H^2\dot{H} f_{TT} = -2\kappa^2 p, \label{F2} &
\end{align}
where the symbol "dot" represents the derivative with respect to cosmic time $t$, $H = \dot{a}(t)/a(t)$ stands for the Hubble parameter. In addition, $\rho$ and $p$ represent the energy density and pressure of the matter content, respectively. Assuming $\kappa^2 = 1$, we can express the aforementioned equations (\ref{F1}) and (\ref{F2}) as
\begin{align}
3H^2 &=\rho+\rho_{DE}, \label{FF1} \\
-2\dot{H} - 3H^2 &=p+p_{DE}. \label{FF2}
\end{align}

Here, the energy density and pressure attributed to DE are established as:
\begin{align}
\rho_{DE} &= -6H^2 f_T - \frac{1}{2} f, \label{rho_DE} \\
p_{DE} &= \frac{1}{2} f + 2\left(\dot{H} + 3H^2\right) f_T + 2H\dot{f}_T. \label{p_DE}
\end{align}

From Eqs. (\ref{rho_DE}) and (\ref{p_DE}), we obtain the expression of the EoS parameter of DE as,
\begin{equation}
    \omega_{DE} =\frac{p_{DE}}{\rho_{DE}}=-1-\frac{4 \left(H \dot{f_T}+\dot{H} f_T\right)}{12 H^2 f_T+f}.
    \label{omega_DE}
\end{equation}

\section{$f(T)$ Cosmology}
\label{sec3}

In the realm of $f(T)$ modified gravity, researchers have extensively explored power-law models to reconstruct and describe various evolution scenarios of the universe. Karami and Abdolmaleki \cite{Karami/2012} examined the generalized second law of thermodynamics within the context of $f(T)$ gravity for two viable models, including the power-law $f(T)$ model. Rezazadeh et al. \cite{Rezazadeh/2016} investigated power-law and intermediate inflationary models within the framework of $f(T)$ gravity. The study explores the implications of $f(T)$ gravity for early universe cosmology. Basilakos \cite{Basilakos/2016} explored the linear growth of structures in the universe within the context of power-law $f(T)$ gravity. Malekjani et al. \cite{Malekjani/2017} analyzed the spherical collapse model and cluster number counts within the framework of power-law $f(T)$ gravity. Boko and Houndjo \cite{Boko/2020} investigated cosmological models incorporating viscous fluids in the framework of $f(T)$ gravity. Their study is centered on characterizing infinite-time singularities within the power-law $f(T)$ model. Recently, Kumar et al. \cite{Kumar/2023} presented new cosmological constraints on $f(T)$ gravity, considering full Planck-CMB and SNeIa data. The study provides updated information on the viability of $f(T)$ gravity as an alternative theory of gravity at cosmological scales. In this paper, we investigate the behavior of the late-time universe by studying the simplest EoS parameterization within the framework of the following power-law $f(T)$ model,
\begin{equation}
    f(T)=\alpha(-T)^n,
    \label{fT}
\end{equation}
where $\alpha$ and $n$ are model parameters. Ideally, we aim to integrate the modified Friedmann equations (\ref{F1})-(\ref{F2}) while considering Eq. (\ref{fT}). However, solving Eqs. (\ref{F1})-(\ref{F2}) simultaneously with Eqs. (\ref{rho_DE}) and (\ref{p_DE}) is challenging due to their complexity. The primary objective of this study is to explore the dynamics of the universe and the characteristics of DE, relying on observational data. As outlined in the introduction, various parameterizations have been proposed, most involving two or more parameters. Mandal et al. \cite{Mandal/2023} explored cosmological observational constraints on the power-law $f(Q)$ type modified gravity theory, using the CPL parametrization form of DE EoS. Arora et al. \cite{EoS1} investigated constraints on the effective EoS in $f(Q, T)$ gravity. The study focuses on understanding the behavior of DE in this modified gravity theory and its implications for cosmology. Koussour and De \cite{EoS2} investigated observational constraints on two cosmological models within the framework of $f(Q)$ theory, using the parametrization form of the EoS parameter as $\omega(z) = -\frac{1}{1+3\beta(1+z)^{3}}$. Myrzakulov et al. \cite{EoS5} examined the non-linear $f(R, L_m)$ DE model, using the same form of the EoS parameter. They present results from Bayesian analysis of cosmic chronometers and Pantheon SNeIa samples, providing a deeper understanding of the behavior of DE in this model. However, using such parameterizations for investigating the dynamics of the universe becomes intricate, as it necessitates introducing additional cosmological parameters, $H_0$ and $n$, to the model. For this reason, a parameterization should have only one parameter. In light of this consideration, we adopt the simplest parametrization of the EoS for DE, as proposed by Gong and Zhang \cite{Gong/2005},
\begin{equation}
    \omega_{DE}(z)=\frac{\omega_0}{1+z},
    \label{EoS_DE}
\end{equation}
where $\omega_0$ represents the current value of the EoS for the DE \cite{Shrivastava/2023}. As $z$ approaches infinity (in the past), $\omega_{DE}(z)=0$ indicates that in the early universe, the DE EoS tends toward zero. At the present redshift ($z=0$), $\omega_{DE}(z)=\omega_0$ represents the current value of the DE EoS. As $z$ approaches $-1$ (in the future), $\omega_{DE}(z)=- \infty $ indicates that in the far future, the DE EoS approaches negative infinity.

From Eqs. \eqref{omega_DE}, \eqref{fT}, and \eqref{EoS_DE}, we have
\begin{equation}
\dot{H}+\frac{3 H^2}{2n}+\frac{3 \omega_{0} H^2}{2n(1+z)}=0.
\end{equation}

Using the relationship $ \frac{1}{H} \frac{d}{dt}= \frac{d}{dln(a)}$ ($a=\frac{1}{1+z}$), we can rephrase the given equation into a first-order differential equation,
\begin{equation}\label{20}
\frac{dH}{dln(a)}+\frac{3 H}{2n}+\frac{3 \omega_{0} H}{2n(1+z)}=0.
\end{equation}

By performing the integration of the above-mentioned equation, we can derive the expression for the Hubble parameter in terms of redshift as,
\begin{equation}\label{Hz}
H(z)=H_{0} (1+z)^{\frac{3}{2n}} \exp \left[ \frac{3\omega _{0}}{2n}\left( \frac{z}{1+z}\right) \right],
\end{equation}
where $H_0$ represents the present value of the Hubble parameter. From Eq. (\ref{Hz}), it is evident that the model parameter $\alpha$ is not directly included in the expression for the Hubble parameter. Therefore, we fix its value to investigate the evolution of DE density and pressure. For our analysis, we set $\alpha=1$.

Moreover, the DE density and pressure can be derived from Eqs. (\ref{rho_DE}), (\ref{p_DE}) and (\ref{Hz}) as
\begin{align}
\rho_{DE}(z) &= 2^{n-1} 3^n \alpha(2 n-1) H_{0}^{2n} (1+z)^{3} \exp \left[ 3\omega _{0}\left( \frac{z}{1+z}\right) \right], \\
p_{DE}(z) &= 2^{n-1} 3^n \alpha(2 n-1) H_{0}^{2n} \frac{\omega_0 (1+z)^{3}}{1+z} \exp \left[ 3\omega _{0}\left( \frac{z}{1+z}\right) \right].
\end{align}

The deceleration parameter assumes a crucial role in characterizing the dynamics of the expansion phase of the universe, and it is defined as
\begin{equation}\label{q}
q=-1-\frac{\dot{H}}{H^2}.
\end{equation}

By using Eq. \eqref{Hz} in Eq. \eqref{q}, we obtain
\begin{equation}\label{qz}
q(z)=-1+\frac{3 (1+z+\omega_{0})}{2 n (1+z)}.
\end{equation}

\section{Constraints from Observations and Methodological Approach} \label{sec4}

In this section, we will perform a statistical analysis using the MCMC approach from the \textit{Python} library \cite{Mackey}. Our objective is to assess the viability of the model by comparing its predictions with various cosmic observations. Specifically, we use datasets from observational Hubble data $H(z)$, SNeIa, and BAO.

\subsection{$H(z)$ datasets}
We use measurements of the Hubble parameter obtained through the differential age method (DAM), commonly referred to as cosmic chronometer data. In this context, we consider a dataset consisting of 31 points compiled in \cite{Moresco/2015}. The $\chi^{2}$ function is defined as follows:
\begin{equation}
\chi _{Hz}^{2}(H_0, \omega_0,n)= \sum_{i=1}^{31} \frac{\left[H(z_{i},H_0,\omega_0,n)-H_{obs}(z_{i})\right]^2}{\sigma^{2}(z_{i})},
\end{equation}
where $H_{obs}$ represents the observed value, and $\sigma(z_{i})$ denotes the observational error associated with each data point.

\subsection{SNeIa datasets}
Since SNe Ia serves as standard candles providing reliable estimates of cosmic distances, they are instrumental in imposing constraints on the DE sector. For our analysis, we leverage the Pantheon compilation consisting of 1048 points distributed across the redshift range $0.01 < z < 2.26$ \cite{Scolnic/2018}. The $\chi^{2}$ function for this dataset is expressed as
\begin{equation}
\chi _{SNeIa}^{2}(H_0, \omega_0,n)=\sum_{i=1}^{1048}\dfrac{\left[ \mu(H_0, \omega_0,n ,z_{i})-\mu
_{obs}(z_{i})\right] ^{2}}{\sigma
^{2}(z_{i})},
\end{equation}%
where $\mu_{obs}$ signifies the observational distance modulus and $\sigma(z_{i})$ represents the observational error corresponding to each data point. In addition, we define $\mu= m_{B}-M_{B}$, where $m_{B}$ denotes the observed apparent magnitude at a given redshift, and $M_{B}$ is the absolute magnitude. The nuisance parameters are obtained using the BBC approach \cite{Kessler/2017}. 

The theoretical distance modulus $\mu(z)$ is
\begin{equation}
\mu(z)= 5 log_{10}\left[\frac{d_{L}(z)}{1 Mpc}\right]+25,.
\end{equation}
where $d_{L}(z)$ denotes the luminosity distance (for a spatially flat universe) and is expressed as
\begin{equation}
    d_{L}(z)= c(1+z) \int_{0}^{z} \frac{dy}{H(y,H_0, \omega_0,n)}.
\end{equation}

Here, $c$ represents the speed of light.

\subsection{BAO datasets}
BAOs refer to pressure waves generated by cosmological perturbations in the baryon-photon plasma during the recombination epoch. These oscillations manifest as distinct peaks on large angular scales. In our analysis, we incorporate BAO measurements obtained from the Six Degree Field Galaxy Survey (6dFGS), Sloan Digital Sky Survey (SDSS), and the LOWZ samples of the Baryon Oscillation Spectroscopic Survey (BOSS) \cite{Blake/2011, Percival/2010}. The mathematical expressions employed for the BAO datasets are:
\begin{eqnarray}
d_{A}(z) &=& c \int_{0}^{z} \frac{dz'}{H(z')},\\
D_{v}(z) &=& \left[\frac{d_{A}(z)^2 c z }{H(z)}\right]^{1/3},\\
\chi_{BAO}^2 &=& X^{T} C^{-1} X, 
\end{eqnarray}
where $d_{A}(z)$ represents the comoving angular diameter distance, $D_{v}(z)$ denotes the dilation scale, and $C$ stands for the covariance matrix \cite{Giostri/2012}.

\subsection{Hz+SNeIa+BAO+CMB datasets}

In addition, we use the CMB data from the most recent and definitive Planck data release \cite{Planck/2018}. Its impact on the likelihood analysis is characterized by the compressed form involving CMB shift parameters \cite{Wang/2007, Wang/2013}:
\begin{eqnarray}
    R &\equiv&\sqrt{\Omega_{m}H_{0}^{2}}r(z_{*})/c, \\
    l_{a} &\equiv& \pi r(z_{*})/r_{s}(z_{*}).
\end{eqnarray}

Here, $r_{s}(z)$ represents the comoving sound horizon at redshift $z$, and $z_{*}$ denotes the redshift to the photon-decoupling surface. In our analysis, we use the Planck data, including temperature and polarization data, and CMB lensing. Here, we rely on estimates derived from the Planck 2018 data, as detailed in Zhai and Wang \cite{Zhai}, using the data vector and covariance data specified in Eq. (31) of the referenced work.

Now, we determine constraints on the parameters of our $f(T)$ cosmological model ($H_0$, $\omega_0$, and $n$) by using the combined $Hz+SNeIa+BAO+CMB$ datasets. This process involved minimizing the total chi-squared function, which incorporates data from the Hz, SNeIa, BAO, and CMB datasets,
\begin{equation}
\chi^{2}_{total} = \chi^{2}_{Hz} + \chi^{2}_{SNeIa}+\chi^{2}_{BAO}+\chi^{2}_{CMB}.
\end{equation}

Then, we generated two-dimensional likelihood contours with 1-$\sigma$ and 2-$\sigma$ errors, corresponding to 68\% and 95\% confidence levels, for the joint ($Hz+SNeIa+BAO+CMB$) datasets. These contours are illustrated in Fig. \ref{Com}. In our joint analysis, we obtained the Hubble constant $H_0$ as $H_0=72.7^{+1.9}_{-1.8}$. In addition, for the parameter $n$, which represents the deviation from the standard model ($n=1$), the constrained value was found to be $n=0.768^{+0.063}_{-0.056}$. Importantly, the Hubble constant values obtained are consistent with various experimental values \cite{Aiola/2020,Wang/2020,Balkenhol/2021,Addison/2021}. The Hubble constant quantifies the present rate of expansion of the universe, a crucial metric in cosmology. Estimating its value and associated uncertainty has been a focal point of research for many years. The $H_0$ tension, referring to the discrepancy between the value of $H_0$ inferred from local measurements within the nearby universe and the value derived from observations of the early universe, remains one of the most significant challenges in modern cosmology \cite{Abdalla/2022}. Local measurements, such as those from the SH0ES project, yield a value of approximately $H_0 = 74.03 \pm 1.42$ $km/s/Mpc$ \cite{Riess:2019cxk}, while observations of the Planck collaboration suggest a lower value, around $H_0 = 67.4 \pm 0.5$ $km/s/Mpc$ km/s/Mpc \cite{Planck/2018}. This leads to a notable tension of 4.4$\sigma$ between the two measurements. Additional discussions regarding the Hubble tension and potential resolutions can be found in Ref. \cite{DiValentino}. In our analysis, when comparing the results of the power-law $f(T)$ model with those of the Planck collaboration \cite{Planck/2018}, we observe a discrepancy of 2.7$\sigma$ in $H_0$ based on the joint analysis of $Hz+SNeIa+BAO+CMB$ data. Further studies addressing the Hubble tension are available in \cite{Yang:2021eud,DiValentino1}.

In addition, for a comparative analysis with the $\Lambda$CDM model, we examined the evolutions of the Hubble parameter $H(z)$ and the distance modulus $\mu(z)$ along with the constraint values of model parameters $H_0$, $\omega_0$, and $n$, using joint datasets. For the standard $\Lambda$CDM model, we use the Hubble parameter expression $H(z) = H_0 \sqrt{\Omega_{m0} (1+z)^3 + \Omega_\Lambda}$, where $\Omega_{m0}$ and $\Omega_\Lambda$ are the present-day matter density parameter and the cosmological constant density parameter, respectively. Using the Planck 2020 results \cite{Planck/2018}, which provide $\Omega_{m0} \approx 0.315$ and $ \Omega_\Lambda \approx 0.685$, these comparisons are illustrated in Figs. \ref{Hubble} and \ref{Mu}, respectively. The findings reveal that our cosmological model aligns well with the observational results in both cases. Furthermore, it is noteworthy that our model closely resembles the profile of the $\Lambda$CDM model. The results indicate that the $f(T)$ gravity model provides a comparable, and in some cases better, fit to the observational data. Despite this resemblance, the $f(T)$ model offers a distinct advantage by predicting a stronger late-time acceleration, as reflected in the present value of the deceleration parameter $q_0$. This feature suggests that the $f(T)$ model could provide a more nuanced understanding of the observed acceleration of the universe.

Furthermore, the EoS parameter $\omega$ is widely recognized for its pivotal role in describing the diverse energy-dominated evolutionary processes of the universe \cite{Myrzakulov/2023}. The current state of the universe can be predicted through either the quintessence phase ($-1<\omega<-\frac{1}{3}$) or the phantom phase ($\omega<-1$). In the context of the present model, we obtain $\omega_0=-0.896^{+0.066}_{-0.063}$, using joint datasets. These results for $\omega_0$ are consistent with findings from various observational studies (refer to \cite{EoS1,EoS2,EoS3,EoS4,EoS5}). It is noteworthy to mention that our $f(T)$ model exhibits quintessence behavior in each data analysis. The obtained constraint values are summarized in Tab. \ref{table1}.

\begin{widetext}

\begin{figure}[h]
\centering
\includegraphics[scale=0.85]{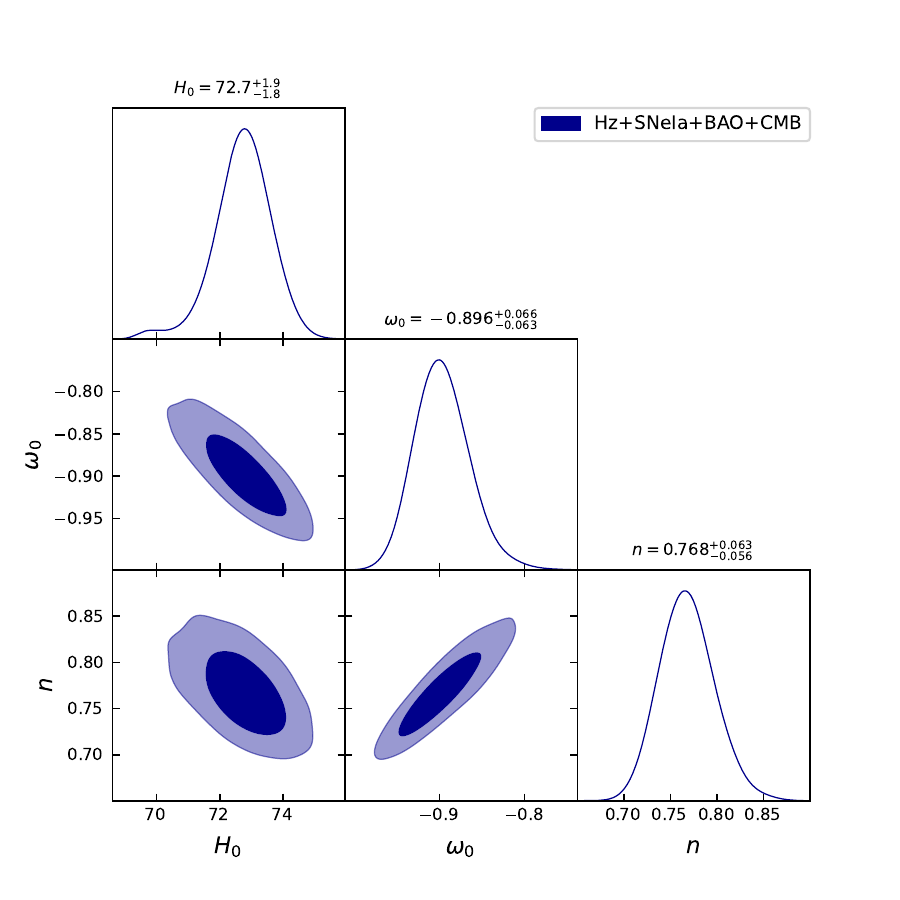}
\caption{Marginalized confidence regions, both one-dimensional and two-dimensional, at 68\% CL and 95\% CL, are presented for the parameters $H_0$, $\omega_0$, and $n$. These results are derived from the joint datasets within the framework of the $f(T)$ gravity model.}
\label{Com}
\end{figure}

\begin{figure}[h]
\centering
\includegraphics[width=18cm,height=5.5cm]{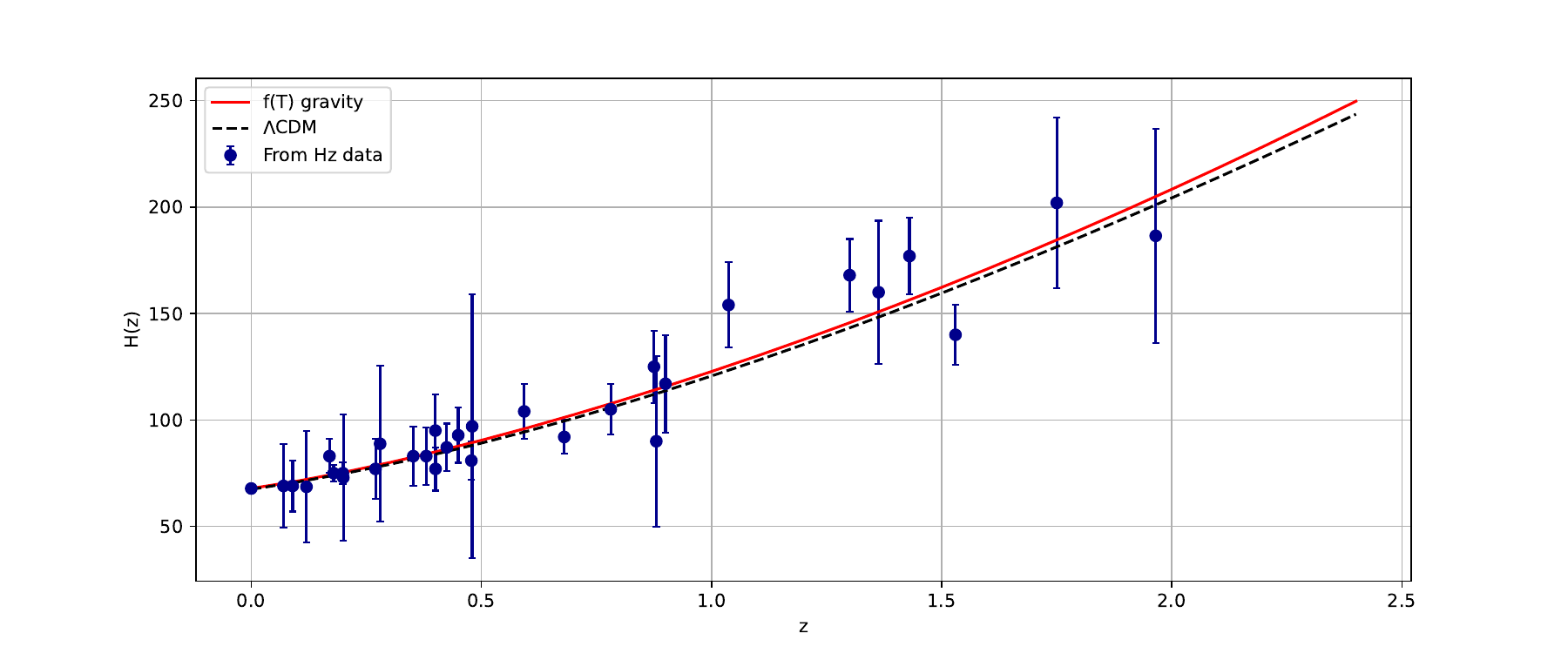}
\caption{The plots depict $H(z)$ against redshift $z$, showing theoretical predictions by the red curve and the $\Lambda$CDM model by the dotted line. The 31 Hubble points, along with their corresponding error bars, are shown as blue dots.}
\label{Hubble}
\end{figure}

\begin{figure}[h]
\centering
\includegraphics[width=18cm,height=5.5cm]{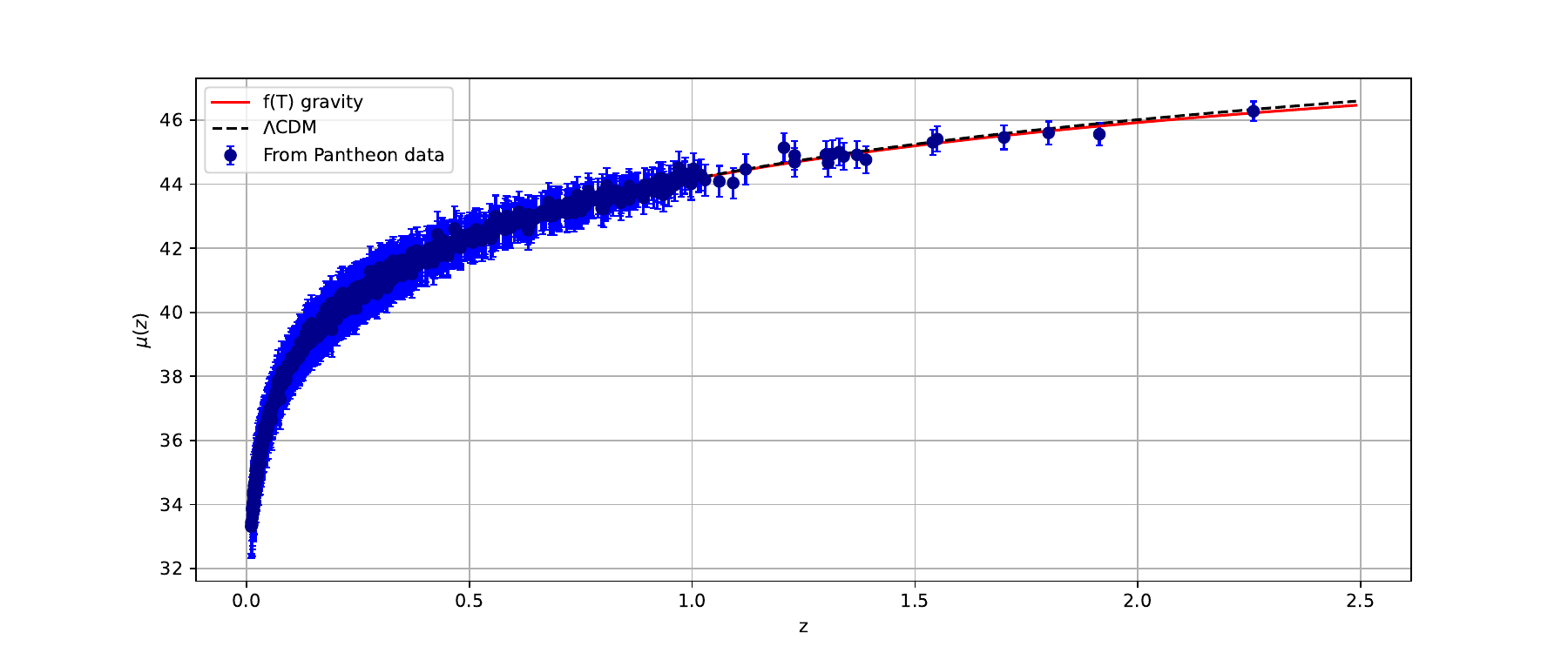}
\caption{The plots depict $\mu(z)$ against redshift $z$, showing theoretical predictions by the red curve and the $\Lambda$CDM model by the dotted line. The 1048 Pantheon points, along with their corresponding error bars, are shown as blue dots.}
\label{Mu}
\end{figure}

\begin{table}[h]
\begin{center}
\begin{tabular}{|l|c|c|c|c|c|}
\hline 
$Datasets$              & $H_{0}$ & $\omega_{0}$ & $n$ & $q_{0}$ & $z_{t}$\\
\hline

$Priors$           & $(60,80)$  & $(-2,2)$ &  $(-10,10)$ & $-$ & $-$\\
\hline
$Hz+SNeIa+BAO+CMB$             & $72.7^{+1.9}_{-1.8}$  & $-0.896^{+0.066}_{-0.063}$ & $0.768^{+0.063}_{-0.056}$ &  $-0.80^{+0.11}_{-0.10}$ & $0.84^{+0.02}_{-0.01}$\\ 
\hline
\end{tabular}
 \caption{Marginalized constrained data for the parameters $H_0$, $\omega_0$, and $n$, along with the corresponding deceleration parameter $q_0$, using the combined $H(z)+SNeIa+BAO+CMB$ datasets at a 68\% confidence level.}
 \label{table1}
\end{center}
\end{table}

\end{widetext}

\section{Dynamic behavior of the cosmos} \label{sec5}

The dynamics behavior of cosmological parameters, comprising the deceleration parameter, DE energy density, and DE pressure, are elucidated below. These descriptions are derived from the constrained values obtained for the model parameters, providing a comprehensive understanding of how these crucial parameters evolve over cosmic time within the framework of the proposed model.

\begin{figure}[h]
\centering
\includegraphics[scale=0.7]{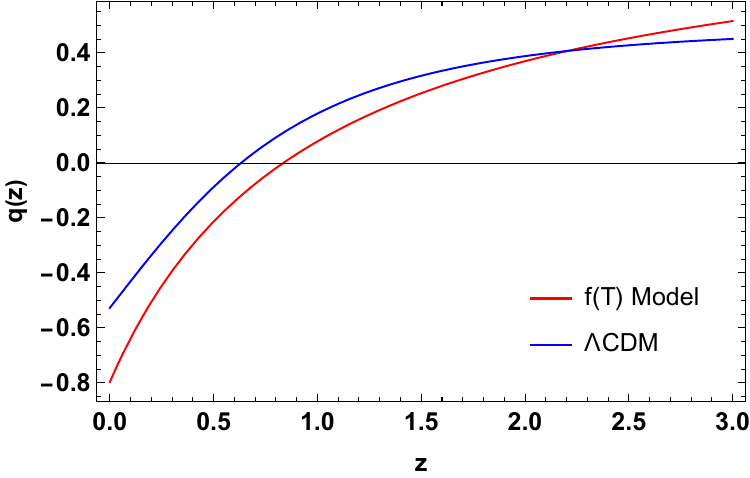}
\caption{Plot of the deceleration parameter versus redshift using the combined $Hz+SNeIa+BAO+CMB$ datasets.}
\label{F_q}
\end{figure}

The deceleration parameter $q$ serves as an indicator of the universe's acceleration or deceleration. Specifically, when $q>0$, the model signifies decelerating expansion, $q=0$ corresponds to a constant rate of expansion, and an accelerating expansion is indicated when $-1<q<0$. Furthermore, the universe exhibits exponential expansion or de Sitter expansion for $q=-1$, while super-exponential expansion occurs for $q<-1$. The sign of $q$ provides valuable insights into the dynamic behavior of the cosmos \cite{Koussour1,Koussour2,Koussour3}. In Fig. \ref{F_q}, it is evident that the universe initiates its history with a decelerating phase and undergoes a transition to an accelerating phase at a specific redshift $z_t$. This observed evolution aligns with the recent behavior of the universe, which has experienced three distinct stages: an early decelerating dominated phase, a subsequent period of accelerating expansion, and finally, a late-time accelerating phase. Also, the figure illustrates that the universe concludes its evolution with a super-exponential expansion $(q<-1)$ at lower redshifts. In our analysis, we plot the deceleration parameter $q(z)$ for both our $f(T)$ gravity model and the standard $\Lambda$CDM model to compare their predictions with observational data. For the $f(T)$ gravity model, the present value of the deceleration parameter is found to be $q_0 \approx -0.80$ (please see Tab. \ref{table1}). This value indicates a strong acceleration of the universe at the present epoch, aligning well with recent observations \cite{Almada/2019,Basilakos/2012,Garza/2019,Jesus/2020}. In contrast, for the standard $\Lambda$CDM model, using the Planck 2018 results \cite{Planck/2018}, we obtain the present value of the deceleration parameter as $q_0 \approx -0.53$. This comparison highlights that our $f(T)$ gravity model predicts a more pronounced acceleration than the $\Lambda$CDM model, suggesting that the former may offer a better explanation for the observed cosmic acceleration.

\begin{figure}[h]
\centering
\includegraphics[scale=0.7]{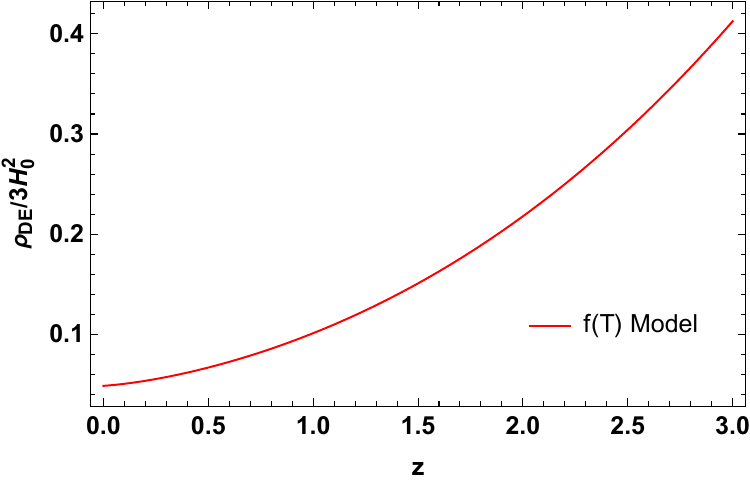}
\caption{Plot of energy density of DE versus redshift using the combined $Hz+SNeIa+BAO+CMB$ datasets.}
\label{F_rho}
\end{figure}

\begin{figure}[h]
\centering
\includegraphics[scale=0.7]{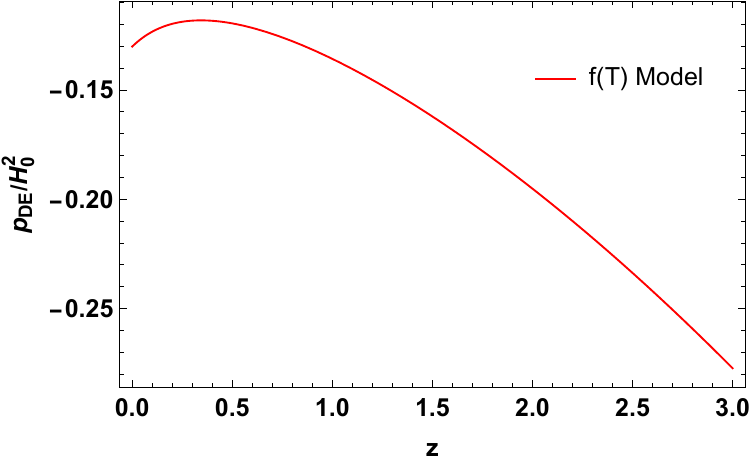}
\caption{Plot of pressure of DE versus redshift using the combined $Hz+SNeIa+BAO+CMB$ datasets.}
\label{F_p}
\end{figure}

Furthermore, the positive behavior of the DE energy density, as depicted in Fig. \ref{F_rho}, suggests a sustained and gradually diminishing influence of DE on cosmic evolution at the present epoch ($z=0$). The figure underscores the enduring nature of DE, indicating its potential to play a significant role in shaping the destiny of the universe over extended cosmic timescales. Also, the behavior of DE pressure, as illustrated in Fig. \ref{F_p}, manifests as a decreasing function of redshift. It initiates from a relatively large negative value and gradually tends towards a smaller negative value at the present epoch ($z=0$). This trend aligns with observational findings, where the negative pressure associated with DE corresponds to the context of the accelerated expansion of the universe. Therefore, the observed behavior of the DE pressure in our model concurs with empirical evidence, further supporting the consistency of our theoretical framework with observed cosmic phenomena.

\section{$Om(z)$ diagnostics} \label{sec6}

The $Om(z)$ diagnostic serves as a valuable tool for categorizing various cosmological models of DE \cite{Sahni}. Distinguished by its simplicity, this diagnostic relies solely on the first-order derivative of the cosmic scale factor. In the case of a spatially flat universe, it is expressed as:
\begin{equation}
Om\left( z\right) =\frac{\left( \frac{H\left( z\right) }{H_{0}}\right) ^{2}-1%
}{\left( 1+z\right) ^{3}-1}.     
\end{equation}

\begin{figure}[h]
\centering
\includegraphics[scale=0.7]{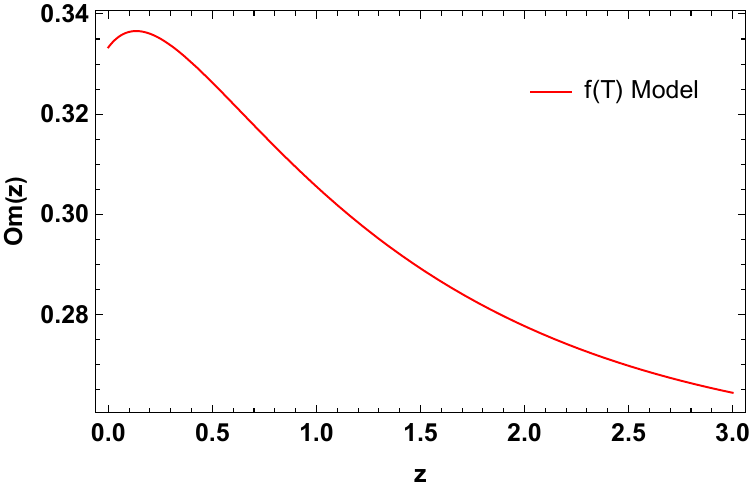}
\caption{Plot of $Om(z)$ diagnostics versus redshift using the combined $Hz+SNeIa+BAO+CMB$ datasets.}
\label{F_Om}
\end{figure}

In this expression, $H_0$ denotes the present value of the Hubble parameter. A negative slope in $Om(z)$ signifies quintessence-type behavior, while a positive slope corresponds to phantom behavior. A constant $Om(z)$ reflects the cosmological constant ($\Lambda$CDM) model. From Fig. \ref{F_Om}, it becomes evident that the $Om(z)$ diagnostic, considering the constrained values of the model parameters, exhibits an initial negative slope that gradually transforms into a positive slope. This observation indicates that our models initially resemble quintessence behavior and eventually transition toward the phantom region. Thus, we can infer that the $Om(z)$ diagnostic behavior aligns with that of the EoS parameter.

\section{Conclusion} \label{sec7}
Motivated by the teleparallel formulation within the scope of GR, our exploration delves into the extension of $f(T)$ gravity by introducing an algebraic function contingent upon the torsion scalar $T$. This article thoroughly investigates the cosmological consequences of the $f(T)$ gravity theory, specifically employing the simplest parametrization for the equation of state parameter of DE, namely $ \omega_{DE}(z)=\frac{\omega_0}{1+z}$. Notably, this parametrization provides $\omega_0$ at the present epoch. Our chosen functional form for $f(T)$ model is $f(T)=\alpha(-T)^n$, representing a power-law dependence on the torsion scalar $T$. Having derived the solution to the modified Friedmann equations, expressed as the Hubble parameter in terms of the redshift $z$, our subsequent step, detailed in Sec. \ref{sec3}, involves leveraging recent observational data. Specifically, We use a joint analysis of $Hz+SNe Ia+BAO+CMB$ to rigorously constrain the unknown parameters in our model ($H_0$, $\omega_0$, $n$). The outcomes of this analysis are depicted in Fig. \ref{Com}, presenting the best-fit values for the model parameters. To gauge the performance of our model against the standard $\Lambda$CDM model, we have compared the $H(z)$ and $\mu(z)$ (distance modulus) in Figs. \ref{Hubble} and \ref{Mu} with cosmic observational data. These visual comparisons provide insights into how well our model aligns with the empirical evidence and demonstrate its efficacy in describing the observed cosmic dynamics.

Within the framework of the current model, the derived values for the EoS parameter at the present epoch are $\omega_0=-0.896^{+0.066}_{-0.063}$, using joint datasets. It is crucial to highlight that our $f(T)$ model consistently demonstrates quintessence behavior throughout our analysis. The model parameters under constraint unveil a diverse array of intriguing cosmological phenomena. One notable observation is the evolution of the deceleration parameter (see Fig. \ref{F_q}), transitioning from a decelerating phase to an accelerating phase, providing a plausible explanation for the late-time universe dynamics. In addition, Figs. \ref{F_rho} and \ref{F_p} depict the behavior of DE components: the DE energy density exhibits a diminishing trend as the universe expands into the distant future. At the same time, the DE pressure showcases a decreasing negative behavior with redshift. Finally, the $Om(z)$ diagnostic presented in Fig. \ref{F_Om} serves as a tool for discerning between different DE models, providing insights into the compatibility of our $f(T)$ model with quintessence scenarios \cite{Koussour4,Koussour5,Koussour6}.

\textbf{Acknowledgment} This research was funded by the Science Committee of the Ministry of Science and Higher Education of the Republic of Kazakhstan (Grant No. AP14972745).

\textbf{Data availability} All data used in this study are cited in the references and were obtained from publicly available sources.

\end{document}